\documentclass[reprint, amsmath, amssymb, aps,superscriptaddress,nofootinbib]{revtex4-1}
\usepackage{graphicx}
\usepackage{dcolumn}
\usepackage{bm}
\usepackage{mathrsfs}
\usepackage{tikz}
\usepackage{hyperref}
\input epsf.tex
\usepackage{amssymb}
\usepackage{amsmath}
\usepackage{amsthm}
\usepackage{amsopn}
\usepackage{cancel}
\usepackage{color}
\usepackage{float}
\usepackage{comment}
\usepackage{multirow}

\usepackage{bbold}

\newcommand{\sm}{\kern0.1em}

\hypersetup{
    colorlinks=true,       
    linkcolor=blue,          
    citecolor=blue,        
    filecolor=magenta,      
    urlcolor=blue           
}

\begin{document}
\title{Arrival Time\textemdash Classical Parameter or Quantum Operator?}
\author{MohammadJavad Kazemi}\email{kazemi.j.m@gmail.com}
\affiliation{Department of Physics, Faculty of Science, University of Qom, Qom, Iran}
\affiliation{Philosophy of Physics Group, Mana Research Institute for Social and Cultural Studies,  Tehran, Iran}
\author{MohammadHossein Barati}\email{m.h.barati1989@gmail.com}
\affiliation{Department of Physics, Faculty of Science, University of Qom, Qom, Iran}
\author{Ghadir Jafari}\email{gh.jafari@cfu.ac.ir}
\affiliation{Department of Physics Education, Farhangian University, P.O. Box 14665-889, Tehran, Iran}
\author{S. Shajidul Haque}\email{shajid.haque@uct.ac.za}
\affiliation{Department of Mathematics \& Applied Mathematics, University of Cape Town, Cape Town, South Africa}
\affiliation{The National Institute for Theoretical and Computational Sciences, \\
Private Bag X1, Matieland, South Africa}
\author{Saurya Das}\email{saurya.das@uleth.ca}
\affiliation{Department of Physics and Astronomy, University of Lethbridge, Lethbridge, Alberta T1K 3M4, Canada}
\begin{abstract} 
The question of how to interpret and compute arrival-time distributions in quantum mechanics remains unsettled, reflecting the longstanding tension between treating time as a quantum observable or as a classical parameter. While recent studies have primarily contrasted arrival-time predictions across different interpretations of quantum mechanics, here we investigate this ambiguity within the standard quantum-mechanical framework itself. Moreover, most previous studies have focused on the single-particle case in the far-field regime, where these distinct approaches yield very similar arrival-time distributions and a semi-classical analysis typically suffices. Recent advances in atom-optics technologies now make it possible to experimentally investigate arrival-time distributions for entangled multi-particle systems in the near-field regime, where a deeper analysis beyond semi-classical approximations is required. Even in the far-field regime, due to quantum non-locality, the semi-classical approximation cannot generally hold in multi-particle systems. Therefore, in this work, two fundamental approaches to the arrival-time problem—namely, the time-parameter and time-operator approaches—are extended to multi-particle systems. Using these extensions, we propose a feasible two-particle arrival-time experiment and numerically evaluate the corresponding joint distributions. Our results reveal regimes in which the two approaches yield inequivalent predictions, highlighting conditions under which experiments could shed new light on distinguishing between competing accounts of time in quantum mechanics. Our findings also provide important insights for the development of quantum technologies that use entanglement in the time domain, including non-local temporal interferometry, temporal ghost imaging, and temporal state tomography in multi-particle systems.
\end{abstract}
\maketitle

\section{Introduction}\label{Intro}
Consider a quantum particle with an initial wave function $\psi$ evolving toward a detector waiting to observe it—arguably the simplest quantum experiment.
When the experiment is repeated with the same initial state, the particle may be detected at different \textit{times}, which can be measured with high precision \cite{korzh2020demonstration,keller2014bose,kurtsiefer19962A}. How can one compute the \textit{temporal} distribution of these detection events?
Remarkably, even a century after the advent of quantum theory, there is still no universally agreed-upon answer to this fundamental question \cite{mielnik1994screen,muga2000arrival,muga2007time,vona2013does,das2021times,das2021questioning,ayatollah2023can,roncallo2023does,goldstein2024spin,drezet2024arrival,maccone2020quantum}.
One of the main obstacles stems from the fact that, according to the principles of quantum theory, the distribution of any observable must be derived from a corresponding self-adjoint operator. Yet, in conventional quantum mechanics, time is not represented by a self-adjoint operator; instead, it is treated as a classical parameter that merely tracks the evolution of the wave function.

Within standard quantum theory, there appear to be only two possible approaches to addressing this arrival time problem. The first treats the arrival time as a quantum observable in the usual sense, which requires introducing an appropriate arrival-time operator \cite{maccone2020quantum,araujo2024space,flores2019quantum,galapon2012only,galapon2005transition,galapon2004confined,kijowski1999comment,grot1996time,art:Delgado-Muga,razavy1969quantum,Aharonov-Bohm}. In the second approach, time is treated merely as a parameter, and without invoking an arrival time operator, the arrival time distribution is derived from other quantum observable distributions. This can be achieved by interpreting the experiment mentioned above as a continuous position measurement---i.e., a dense sequence of position measurements \cite{allcock1969timeII,mielnik1994screen,dhar2015quantum,dubey2021quantum,friedman2017quantum,thiel2018first}.

Each approach faces significant challenges. On the one hand, as Pauli pointed out, no self-adjoint time operator satisfies $[\hat{H}, \hat{t}] = i\hbar$ when the Hamiltonian is bounded from below \cite{muga2007time,Pauli1958Encyclopedia}. On the other hand, without treating time as an operator, the most natural model of a \textit{continuously waiting detector} forbids detection altogether, a manifestation of the quantum Zeno effect \cite{misra1977zeno,allcock1969timeII,mielnik1994screen,thiel2018first,dhar2015quantum}.
Several innovative approaches have been proposed to address these challenges, employing mathematical techniques and taking into account the specifics of real (rather than idealized) experiments \cite{grot1996time,galapon2005transition,hegerfeldt2010manufacturing,echanobe2008disclosing,dubey2021quantum} (see also Sec.~\ref{Theoretical-Frameworks}). 

Most previous studies of the arrival-time problem have focused on simple cases, such as a free single particle in one-dimension, and have not yet been fully extended to more complex situations\footnote{To the best of our knowledge, the time-parameter approach has not yet been investigated for multi-particle systems. Furthermore, while there is an interesting multi-particle extension of the time-operator approach \cite{baute2002quantum}, it differs from the scenario under consideration here. Whereas the work in \cite{baute2002quantum} addresses the arrival-time distribution of particles at a single detector (yielding a single-time distribution), our focus is on a more general configuration involving two spatially separated detectors, which necessitates a multi-time arrival distribution. Such an extension is required for investigating nonlocal effects in the time domain, e.g.,  non-local temporal interference \cite{ayatollah2024non}. It is worth noting that analogous extensions to multi-particle systems have recently been pursued within alternative interpretations of quantum mechanics, most notably Bohmian mechanics \cite{kazemi2023detection,tumulka2022detection,durr2004exit}.
In addition to employing the standard interpretation rather than alternative interpretations, we have taken into account key physical effects—such as particle indistinguishability and detector back-action—that distinguish our work from these studies.
}. 
Although approaches mentioned above are fundamentally distinct, their predictions often coincide for single-particle case under common experimental conditions—particularly in the regime where the detector is placed in the far-field (or scattering) limit. In fact, in this regime, a semi-classical approximation\footnote{In this approximation,
particles are assumed to follow classical trajectories, and the arrival time distribution is estimated from the quantum momentum distribution \cite{vona2013does,vona2015role,Das2019Exotic,Shucker1980,wolf2000ion}. However, it should be noted that due to the quantum backflow effect \cite{bracken1994probability}, even in free
space, the time evolution of the quantum position
probability density is not consistent with the underlying
uniform motion, especially in near-field regime \cite{Hofmann2017}.
}
is typically sufficient to estimate the arrival time distribution \cite{vona2013does,vona2015role,das2022double,wolf2000ion,kurtsiefer1995time}.
However, thanks to recent advances in atom optics—particularly the development of fast single-atom detectors \cite{khakimov2016ghost,keller2014bose,allemand2024tomography}, and  generation and optical manipulation of entangled states \cite{bergschneider2019experimental,omran2019generation,lopes2015atomic}—it is now possible to explore entangled multi-particle systems in the near-field regime, where the semiclassical approximation breaks down, and a deeper analysis is required  \cite{vona2013does,ayatollah2023can}. Moreover, even in the far-field regime, due to quantum nonlocality, the semi-classical approximation cannot generally hold in multi-particle systems---see also \cite{ayatollah2024non,franson1989bell,das2021questioning}.
Furthermore, in multi-partite arrival-time experiments, particle detections induce wave function collapses throughout the experiment, affecting the joint arrival-time statistics. Consequently, the fundamentally different post-measurement state updates in the aforementioned approaches may amplify differences in multi-particle arrival-time predictions \cite{anastopoulos2017time,kazemi2023detection}.

\begin{figure*}
\centering
\begin{tikzpicture}
\draw (-1,0) node[above right]{\includegraphics[width=18cm]{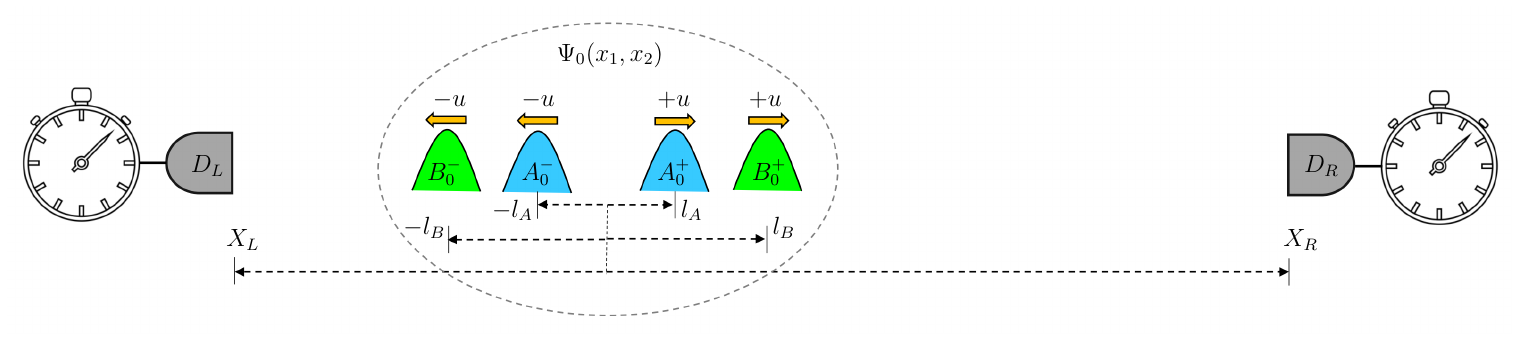}};
\end{tikzpicture}
\caption{\textbf{Schematic diagram of the proposed setup.} In this setup, two entangled particles are prepared in the initial entangled state $\Psi_0(x_1,x_2)$, composed of four Gaussian wave packets, as described in Eq. (\ref{Initial-wave-function}). The left detector $D_L$ and the right detector $D_R$ are waiting to measure the arrival times of the left-moving and the right-moving particles, respectively.}\label{fig1}
\end{figure*}

Accordingly, in this work, we extend two mentioned fundamental approaches to entangled multi-particle systems. Based on this extension, we propose and numerically simulate a simple two-particle arrival-time experiment that is feasible with current atom-interferometry technologies (see Fig.\ref{fig1} and Section \ref{Results} for details). Our numerical results demonstrate that the arrival-time distributions predicted by the two approaches differ sufficiently to be experimentally distinguished. These findings not only shed new light on the foundations of quantum theory but also provide a more precise understanding of quantum phenomena in the time domain, which is crucial for progress in various quantum technologies that use entanglement in time domain, including non-local temporal interferometry \cite{ayatollah2024non,szriftgiser1996atomic}, temporal ghost imaging \cite{khakimov2016ghost,ryczkowski2016ghost}, temporal state tomography \cite{brown2023time,tenart2021observation}, and others \cite{geiger2018proposal,lopes2015atomic}.

The rest of the paper is organized as follows. In Section \ref{Theoretical-Frameworks}, we briefly review the two aforementioned arrival-time approaches and extend them to two-particle systems. In Section \ref{Results}, we present and compare the simulation results for the two-particle distributions based on these approaches. Finally, in Section \ref{Summary}, we summarize the findings and discuss future prospects.

\section{Theoretical Frameworks}\label{Theoretical-Frameworks}
In this section, we review two main approaches to the arrival-time problem within the standard quantum theory framework—the \textit{time-operator}  and the \textit{time-parameter} approaches—and extend them to entangled multi-particle systems, with particular emphasis on the two-particle case. 
\subsection{Arrival Time Operator}
The earliest attempts to address the arrival-time problem within the framework of standard quantum theory date back to the early 1960s, with pioneering works by Aharonov and Bohm \cite{aharonov1961time} and, independently, by Paul \cite{paul1962quantenmechanische}. These studies began with a symmetric quantization of the classical expression for the arrival time. In the one-dimensional case, the classical arrival time of a freely moving particle of mass \( m \), initially located at \( x_0 \) with momentum \( p_0 \), as it travels to a point \(x=L \), is given by
\begin{equation}
  t =m(L-x_0)/p_0,
 	\label{SC0}
\end{equation}
The symmetric quantization of this classical expression yields the following arrival time operator \cite{muga2000arrival}:
\begin{equation}
     \hat{t}_{L}=mL \, \hat{p}^{\,-1}-\frac{m}{2}\,(\hat{p}^{\,-1}\,\hat{x}\,+\,\hat{x}\,\hat{p}^{\,-1}),   
     \label{ABquant}
\end{equation}
Here, \(\hat{x}\) and \(\hat{p} = -i \, \partial / \partial x\) represent the standard position and momentum operators, respectively, while \(\hat{t}_{L}\) is referred to as the Aharonov-Bohm time operator. This operator satisfies the canonical commutation relation with the free Hamiltonian operator, \([\hat{t}_{L}, \hat{p}^2/2m] = i\hbar\), which has been used to derive the energy-time uncertainty relation\footnote{However, this uncertainty relation is not well-defined for all conceivable wave functions, because for many of them, the corresponding arrival time distribution decays too slowly to have finite first and second moment \cite{das2021times}.} 
\cite{giannitrapani1997positive,art:Grot-Rovelli-Tate}.
However, while \(\hat{t}_{L}\) is Hermitian (or symmetric, as referred to in mathematical contexts), it is not a self-adjoint operator \cite{egusquiza1999free}—a result consistent with Pauli's theorem \cite{Pauli1958Encyclopedia}.

The non-self-adjoint nature of $\hat{t}_{AB}$ arises from the singularity at $p = 0$ in its momentum representation,  
$\hat{t}_{L} \to mLp^{-1}-(i\hbar m/2)\,(2p^{-1}\partial_p-p^{-2})$~\cite{egusquiza1999free}.  
Grot, Rovelli, and Tate addressed this singularity in Eq.~\eqref{ABquant} by introducing a symmetric regularization and derived the corresponding arrival-time probability density using the standard Born rule~\cite{art:Grot-Rovelli-Tate}, expressed as~\cite{giannitrapani1997positive,egusquiza1999free}:
\begin{equation}
\Pi(t|x=L) = |\langle t|\psi_0\rangle|^2,
\end{equation}
where $|t\rangle$ denotes the eigenstate of the arrival-time operator $\hat{t}_{AB}$ corresponding to the temporal eigenvalue $t$.  
In the momentum representation, the arrival-time probability density takes the form
\begin{equation}\label{ABdis}
\Pi(t|x\!=\!L) \!=\! \frac{1}{2\pi \hbar} 
\sum_{\alpha=\pm} 
\left| 
\int_{-\infty}^{\infty} \!\!\!dp\, 
\theta(\alpha p)\,
\sqrt{\frac{|p|}{m}}\,
\tilde{\psi}_t(p)\,
e^{\frac{iLp}{\hbar}} 
\right|^2\!\!,
\end{equation}
where $\theta(\cdot)$ is the Heaviside step function, and $\tilde{\psi}_t(p)$ is the momentum-space wave function related to the initial state by
$
\tilde{\psi}_t(p) = \tilde{\psi}_0(p) 
\exp\!(-it p^2/2m\hbar).
$
Extensions of equations \eqref{ABquant} and \eqref{ABdis} to include interaction potentials have been explored in various studies \cite{galapon2004shouldn,galapon2004confined,galapon2005transition,galapon2009theory,flores2019quantum,galapon2012only,hegerfeldt2010manufacturing,das2021times}. These works demonstrate that the non-self-adjoint nature of the free arrival-time operator can be resolved through spatial confinement as well \cite{galapon2002self,galapon2004confined}. Specifically, the Eq.~(\ref{ABdis}) can be obtained as the limiting case of the arrival-time distribution in a confining box as the box length approaches infinity \cite{galapon2005transition}.  Additionally, recent studies have derived the distribution \eqref{ABdis} from a space-time-symmetric extension of non-relativistic quantum mechanics \cite{dias2017space,araujo2024space}. The distribution (\ref{ABdis}) and its generalizations in the presence of interaction potentials
have been referred to as the "standard arrival-time distribution" by several authors \cite{muga2007time,egusquiza2003comment,leavens2005spatial,leavens2007peculiar,das2021times}\footnote{It is worth noting that certain paradoxical properties of the standard arrival-time distribution have been reported in the literature and long debated \cite{leavens2002standard,egusquiza2003comment,leavens2005reply,sombillo2016particle,das2021times}. 
Our objective in the present work, however, is not to settle these theoretical disputes; instead, we emphasize the possibility of validating this distribution through direct comparison with experimental results.}.

We now generalize this standard arrival-time distribution to the two-particle case. 
Fig.~\ref{fig1} illustrates a schematic diagram of a typical experimental setup designed to measure the arrival times of two entangled particles. 
The setup consists of a source that produces pairs of entangled particles, such as helium atoms. 
The particles propagate in opposite directions toward two detectors located at distances $X_L$ and $X_R$. 
These detectors are configured to precisely record the arrival times of the particles, denoted by $t_L$ and $t_R$---the instants at which the left and right
detectors click, respectively.
The goal is to compute the joint two-particle arrival-time probability density, $\Pi(t_L,t_R)$, based on the particles’ initial state $|\Psi_0\rangle$.
In this regard, we generalize the standard single-particle
arrival-time probability density to the
case of two identical particles, in the usual manner, as follows:
\begin{equation}
\Pi(t_L,t_R \,|X_L,X_R) = \left|\langle t_L,t_R|\Psi_0\rangle\right|^2,
\end{equation}
where, for bosons, $|t_L,t_R\rangle$ is constructed as the symmetric tensor product of  the single-particle arrival-time eigenstates,
$$
|t_L,t_R\rangle = |t_L\rangle_{X_L}^{(1)} \otimes |t_R\rangle_{X_R}^{(2)}+ |t_R\rangle_{X_R}^{(1)} \otimes |t_L\rangle_{X_L}^{(2)}.
$$
For fermions, the plus sign must be replaced by a minus sign (antisymmetrization), as required by the exchange statistics.
Accordingly, in the momentum representation, the two-particle arrival-time probability density is given by
\begin{widetext}
\begin{eqnarray}\label{OTA-P-rep}
\Pi(t_L,t_R|X_L,X_R)
\!=\!\!\!
\!\sum_{\alpha_1,\alpha_1=\pm}\! \frac{1}{(2\pi\hbar)^2} \left| \int  dp_1 dp_2 \tilde{\Psi}_0(p_1,p_2)
\theta(\alpha_1 p_1)\theta(\alpha_2 p_2)\sqrt{\frac{p_1}{m}}\sqrt{\frac{p_2}{m}}e^{-\frac{i}{2m\hbar}(p_1^2t_L+p_2^2t_R)}e^{\frac{i}{\hbar}(p_1X_L+p_2X_R)}\right|^2\!\!. \ \ \ \ 
\end{eqnarray}
\end{widetext}
Note that, the wave function must be symmetric,
$\tilde{\Psi}_0(p_1,p_2)=\tilde{\Psi}_0(p_2,p_1)$, for bosons, and antisymmetric,
$\tilde{\Psi}_0(p_1,p_2)=-\tilde{\Psi}_0(p_2,p_1)$, for fermions.

\subsection{Arrival time experiment as a sequence of position measurements}

As previously discussed in the Introduction \ref{Intro}, by treating time as a classical parameter rather than as a quantum observable, the most natural approach to modeling arrival-time experiments is to describe them as a sequence of position measurements. For a single-particle arrival-time experiment, the procedure is as follows: a particle is released from a specified region at time $t = 0$, evolving unitarily toward the detector, and within the detection region, repeated instantaneous projective measurements are conducted at regular time intervals $\tau$ to ascertain whether the particle has arrived. The experiment is finished as soon as the particle is detected for the first time. If the particle is detected at the $n$th measurement, the particle’s time of arrival into the detection region is taken to be $t = n\tau$ (or, more precisely, between $t-\tau$ and $t$).
In a few words, the detector outputs can be represented as a binary string of length $n$, describing detection success: ``no, no, $\ldots$, no, yes''. It is important to note that the probability of first detection at the $n$th measurement is not the same as the probability of detection in 
$t=n\tau$
given no previous measurements. This is because, after each \textit{null} measurement---i.e., a measurement without particle detection---the wave function is projected outside the detection region, according to the standard von Neumann measurement postulate. Consequently, the unitary evolution of the wave function is interrupted by a sequence of projections made at regular time intervals $\tau$. Therefore, using the Born rule, the probability of the first arrival at the $n$th measurement is given by~\cite{allcock1969timeII,porras2014quantum,dhar2015quantum,friedman2017quantum,thiel2018first}:
\begin{equation}\label{one-part-time-param}
\tau\Pi_n=(1-\mathcal{P}_1)(1-\mathcal{P}_2)\cdots(1-\mathcal{P}_{n-1})\mathcal{P}_{n},
\end{equation}
where $\mathcal{P}_{k}$ is defined as
\begin{equation}
\mathcal{P}_{k}=\int_{\mathcal{D}} |\psi_{k}(x)|^2\, dx,
\end{equation}
in which $\mathcal{D}$ is the active domain of the detector, and the wave function at $t=k\tau$ is given by
\begin{equation}\label{one-particle-evolution}
\psi_{k}(x)=N_k\langle x|e^{-i\hat{H}\tau/\hbar} \left(\hat{\bar{P}}e^{-i\hat{H}\tau/\hbar}\right)^{k-1}|\psi_{0}\rangle,
\end{equation}
where $\hat{H}$ is the (free) Hamiltonian, $\hat{\bar{P}}$ is the projection operator corresponding to the null measurements, defined as
$
\hat{\bar{P}}=\mathbb{1}-\int_{\mathcal{D}} dx |x\rangle\langle x|,
$
and $N_k$ is the normalization constant---after each projection, the wave function must be re-normalized. Beyond the arrival-time problem, such non-unitary quantum dynamics of a system subjected to repeated monitoring have garnered significant attention recently, largely due to the growing interest in quantum information (see, e.g., \cite{burgarth2020quantum,thiel2020uncertainty,friedman2017quantum,thiel2018spectral}).

Continuous monitoring of this kind—where the time interval between measurements is taken to zero—leads to a surprising outcome: the quantum particle is never detected, a phenomenon known as the quantum Zeno effect \cite{misra1977zeno,allcock1969timeIII,thiel2018first}.
This phenomenon arises from the general properties of quantum dynamics, which imply a quadratic behavior in the survival probability at short time scales \cite{facchi2002quantum,home1997conceptual,nakazato1996temporal}; when $\tau \to 0$, the squared norms of the
wave-function in the detection region
tend to zero as quickly as $o(\tau^2)$, and so the total probability of particle detection in a fixed time interval $[0, T]$ vanishes as $o(\tau)$ \cite{mielnik1994screen}. In fact, the limit $\tau \to 0$ represents an infinitely fast and sensitive detector, and the quantum Zeno effect demonstrates that such a perfect detection screen, if it existed, could not work---Its unlimited sensitivity would cause the complete reflection of the wave packet rather than its detection. In contrast, it appears that any real detection screens possess finite sensitivity and a corresponding finite "awareness" time $\tau$. This finite sensitivity results in a small but non-zero probability that (wave function of) the particle may pass through the screen before being detected, thereby ensuring a non-zero detection probability.

Let us extend this time-parameter approach to the two-particle case. While the generalization follows naturally from the principles of quantum theory, it is not trivial. The total wave function of the particles evolves according to the Schr\"odinger equation in the two-particle space. However, when a measurement is performed on one part of the system, the wave function must be projected onto the subspace corresponding to the measurement outcome. In our setup, two detectors are waiting for particle arrivals. This \textit{waiting} is modeled as a sequence of measurements performed at time intervals of \(\tau\). At each time step, if one or both detectors register (or fail to register) a particle, the total wave function is projected onto a subspace consistent with the measurement result. Suppose that during the first \(n\) time steps, neither detector detects a particle. In each of these steps, the wave function is projected onto the subspace corresponding to no detection. When, at a later time step, one detector registers a particle while the other does not, the wave function is then projected onto the corresponding subspace; specifically, when the first particle is \textit{absorbed} by a macroscopic detector, the two-particle wave function effectively reduces to a single-particle wave function.  
 Therefore, the two-particle joint arrival time distribution, i.e., the joint probability that one particle is detected by the left detector at a time $t_L \in [(n\!-\!1)\tau, n\tau]$ and the other particle is detected by the right detector at a time $t_R \in [(m\!-\!1)\tau, m\tau]$, is given by\footnote{In principle, both particles may be detected by the same detector. However, we do not consider this case in the present work, since, for the initial wave function considered here, the probability of such an event is very small; see Fig. \ref{fig1} and Sec. \ref{Results}.}
\begin{widetext}
\begin{eqnarray}\label{Joint-TPA}
\tau^2\Pi_{n,m}=
\begin{cases}
(\bar{\mathcal{P}}_1...\bar{\mathcal{P}}_{n-1})\tilde{\mathcal{P}}^{(L)}_{n}(\bar{\mathcal{P}}^{(R)}_{n,n+1}...\bar{\mathcal{P}}^{(R)}_{n,m-1})\mathcal{P}^{(R)}_{n,m} &\ \ n<m \\
\\
(\bar{\mathcal{P}}_1...\bar{\mathcal{P}}_{n-1})\mathcal{P}_n  &\ \ n=m \\ \\
(\bar{\mathcal{P}}_1...\bar{\mathcal{P}}_{m-1})\tilde{\mathcal{P}}^{(R)}_{m}(\bar{\mathcal{P}}^{(L)}_{m,m+1}...\bar{\mathcal{P}}^{(L)}_{m,n-1})\mathcal{P}^{(L)}_{m,n} &\ \ n>m
\end{cases}
\end{eqnarray}
\end{widetext}
where, using the Born rule, the probabilities are given by
\begin{eqnarray}
\mathcal{P}_k\!&=&\!2 \int_{\mathcal{D}_L}dx_1\int_{\mathcal{D}_R}dx_2 \ |\langle x_1, x_2|\Psi_{k}\rangle|^2,\label{P_k}\\
\bar{\mathcal{P}}_k\!&=&\!2 \int_{\bar{\mathcal{D}}_L}dx_1\int_{\bar{\mathcal{D}}_R}dx_2 \ |\langle x_1, x_2|\Psi_{k}\rangle|^2,\label{Pbar_k}\\
\tilde{\mathcal{P}}^{(L)}_{k}\!&=&\!2 \int_{\mathcal{D}_L}dx_1\int_{\bar{\mathcal{D}}_R}dx_2 \ | \langle x_1, x_2|\Psi_{k}\rangle|^2,\label{Ptilde1_k}\\
\tilde{\mathcal{P}}^{(R)}_{k}\!&=&\!2 \int_{\bar{\mathcal{D}}_L}dx_1\int_{\mathcal{D}_R}dx_2 \ | \langle x_1, x_2|\Psi_{k}\rangle|^2,\label{Ptilde2_k}\\
\mathcal{P}^{(L)}_{k,k+l}\!&=&\!\int_{\mathcal{D}_L}dx \ \langle x|\hat{\rho}^{(L)}_{k,l}|x\rangle,\\
\mathcal{P}^{(R)}_{k,k+l}\!&=&\!\int_{\mathcal{D}_R}dx \ \langle x|\hat{\rho}^{(R)}_{k,l}|x\rangle,\\
\bar{\mathcal{P}}^{(L)}_{k,k+l}\!&=&\!\int_{\bar{\mathcal{D}}_L}dx \ \langle x|\hat{\rho}^{(L)}_{k,l}|x\rangle,\\
\bar{\mathcal{P}}^{(R)}_{k,k+l}\!&=&\!\int_{\bar{\mathcal{D}}_R}dx \ \langle x|\hat{\rho}^{(R)}_{k,l}|x\rangle.
\end{eqnarray}
in which $\mathcal{D}_L$ and $\mathcal{D}_R$ represent the active domain of the left and right detectors, respectively. In our setup, we employ open-ended intervals to describe the active detection regions\footnote{Under the considered experimental scenario, this is a well-justified approximation due to the significant disparity in length scales. The wave function penetration depth into the detector’s active region is sub-micrometer, which is substantially smaller than the active volume thickness—typically exceeding a millimeter in usual microchannel plate (MCP) detectors \cite{keller2014bose,hoendervanger2013influence}.}, i.e., $\mathcal{D}_L=(-\infty,X_L]$, $\mathcal{D}_R=[X_R,\infty)$, $\bar{\mathcal{D}}_{L}=\mathbb{R}-\mathcal{D}_{L}$, and $\bar{\mathcal{D}}_{R}=\mathbb{R}-\mathcal{D}_{R}$. The factor of 2 in Eqs. (\ref{P_k}-\ref{Ptilde2_k}) arises from particles indistinguishability.
The two-particle wave function after no detection of either particle at time step \(k\) is given by
\begin{equation}
|\Psi_{k}\rangle = N_k e^{-i\hat{\mathsf{H}}\tau/\hbar}\left(\hat{\bar{P}}e^{-i\hat{\mathsf{H}}\tau/\hbar}\right)^{k-1}|\Psi_{0}\rangle,
\end{equation}
where $N_k$ is the normalization constant, $\hat{\mathsf{H}}$ is the total Hamiltonian of the two-particle system, and $\hat{\bar{P}}$ is the symmetric two-particle projection operators, which is given by:
\begin{eqnarray}
\hat{\bar{P}}\!  =\! \frac{1}{2} \! \int_{\bar{\mathcal{D}}_L} \! \! \! \! dx_1\! \! \int_{\bar{\mathcal{D}}_R} \! \! \! \! dx_2 \big(|x_1,x_2\rangle\langle x_1,x_2|\!+\!|x_2,x_1\rangle\langle x_2,x_1|\big)\! . \ \
\end{eqnarray}
The one-particle quantum states after a particle is detected at time step \(k\) by one detector while the other detector has not yet recorded a detection event at time step \(k+l\), are given by following density matrices,
\begin{eqnarray}
\hat{\rho}^{(R)}_{k,l} &\!\!\!\!=\!\!\!\! & N_{kl}^{(R)} e^{-\frac{i\tau}{\hbar}\hat{H}}\!\!\left(\!\hat{\bar{P}}_{R}e^{-\frac{i\tau}{\hbar}\hat{H}}\!\right)^{l-1}\!\!\!\hat{\rho}^{(R)}_{k,0}\!\left(\! e^{\frac{i\tau}{\hbar}\hat{H}}\hat{\bar{P}}_{R}\!\right)^{l-1}\!\!\! e^{\frac{i\tau}{\hbar}\hat{H}}\!,\ \ \ \ \  \label{psi-ket-R}\\
\hat{\rho}^{(L)}_{k,l} &\!\!\!\!=\!\!\!\! & N_{kl}^{(L)} e^{-\frac{i\tau}{\hbar}\hat{H}}\!\!\left(\!\hat{\bar{P}}_{L}e^{-\frac{i\tau}{\hbar}\hat{H}}\!\right)^{l-1}\!\!\!\hat{\rho}^{(L)}_{k,0}\!\left(\! e^{\frac{i\tau}{\hbar}\hat{H}}\hat{\bar{P}}_{L}\!\right)^{l-1}\!\!\! e^{\frac{i\tau}{\hbar}\hat{H}}\!,\ \ \ \ \  \label{psi-ket-L}
\end{eqnarray}
where  $\hat{H}$ are the single-particle Hamiltonian, $N^{(L)}_{kl}$ and $N^{(R)}_{kl}$ is normalization constants, $\hat{\bar{P}}_{L}$, and  $\hat{\bar{P}}_{R}$ are the  one-particle projection operators, which are given by
\begin{eqnarray}
\hat{\bar{P}}_{L}\!\!  &=&\!\! \frac{1}{2}  \int_{\bar{\mathcal{D}}_L}\! dx |x\rangle\langle x|, \\
\hat{\bar{P}}_{R}\!\!  &=&\!\! \frac{1}{2}  \int_{\bar{\mathcal{D}}_R}\! dx |x\rangle\langle x|.
\end{eqnarray}
The initial one-particle density matrices are given by
\begin{eqnarray}
\hat{\rho}^{(R)}_{k,0}\!\!  &=&\! \! N_{k,0}^{(R)} \! \! \int \! \! dx \! \int \! \! dx' \!\! \int_{D_L} \! \! \! d\xi \ \Psi_k(x,\xi)\Psi^*_k(x',\xi)|x\rangle \langle x'|, \ \ \ \ \ \ \\
\hat{\rho}^{(L)}_{k,0}\!\!  &=&\! \! N_{k,0}^{(L)} \! \! \int \! \! dx \! \int \! \! dx' \! \! \int_{D_R} \! \! \! d\xi \ \Psi_k(x,\xi)\Psi^*_k(x',\xi)|x\rangle \langle x'|, \ \ \ \ \ \ 
\end{eqnarray}
in which $\Psi_k(x,\xi)=\langle x,\xi|\Psi_k\rangle$. Note that, If the detector operates sufficiently rapidly such that the penetration of the wave function into the detection region between successive measurements is small, the state of the remaining particle may, to a good approximation, be regarded as a pure state and written as follows:
\begin{eqnarray}
|\psi^{(R)}_{k,l}\rangle &\!\!=\!\! & N_{kl}^{(R)} e^{-i\hat{H}\tau/\hbar}\left(\hat{\bar{P}}_{R}e^{-i\hat{H}\tau/\hbar}\right)^{l-1}|\psi^{(R)}_{k,0}\rangle,\ \ \ \ \ \ \label{psi-ket-R}\\
|\psi^{(L)}_{k,l}\rangle &\!\!=\!\! & N_{kl}^{(L)} e^{-i\hat{H}\tau/\hbar}\left(\hat{\bar{P}}_{L}e^{-i\hat{H}\tau/\hbar}\right)^{l-1}|\psi^{(L)}_{k,0}\rangle,\ \ \ \ \ \ \label{psi-ket-L}
\end{eqnarray}
in which the initial single-particle states, following the detection of the other particle at the left and right detectors, are given by 
\begin{eqnarray}
|\psi^{(R)}_{k,0}\rangle \!\!=\!\! \int dx \Psi_k(x,X_{L}) |x\rangle,\ \ \ \ \ \ \label{psi-ket-R0}
\end{eqnarray}
and
\begin{eqnarray}
|\psi^{(L)}_{k,0}\rangle \!\!=\!\! \int dx \Psi_k(x,X_{R}) |x\rangle,\ \ \ \ \ \ \label{psi-ket-L0}
\end{eqnarray}
respectively \cite{tumulka2022detection}. The parameters characterizing the proposed experimental setup considered below lie within the regime of validity of this approximation.

\section{Numerical Simulation and Results}\label{Results}
In this section, we numerically study the differences between the two mentioned arrival-time approaches—namely, the time-operator and time-parameter frameworks—in a minimal two-particle arrival-time experiment. The experimental configuration is schematically illustrated in Fig.~(\ref{fig1}). In this setup, a pair of entangled particles is prepared near the origin with an initial wave function composed of two double wave packets—one propagating to the right and the other to the left. On the left side at $x = X_L$ and on the right side at $x = X_R$, there are two particle detectors waiting to detect the particles and record the detection times. By repeating the experiment many times, the joint two-particle arrival-time distribution can be reconstructed from the ensemble of recorded arrival times.

The initial wave function can be considered as follows:
\begin{eqnarray}\label{Initial-wave-function}
\!\!\Psi_{0}(x_1,x_2)\! &=& \! N_0\!\left( A_0^+ (x_1)A_0^-(x_2)\!+\!B_0^+(x_1)B_0^-(x_2) \right. \ \ \ \ \nonumber \\ 
\! &+& \left.\!  1  \leftrightarrow  2 \right)\!,
\end{eqnarray}
where $N_0$ is normalization constant,    
\begin{eqnarray}\label{fGG-def}
A_0^{\pm}(x)&=&G(x;\sigma_0,\pm l_A, \pm u),\\
B_0^{\pm}(x)&=&G(x;\sigma_0,\pm l_B, \pm u),
\end{eqnarray}
and $G$ is a Gaussian wave function as follows:
\begin{equation}
G(x;\sigma, l,  u)=Ne^{-(x-l)^2/4 \sigma^2+ i m u (x-l)/\hbar}.
\end{equation}
The $1 \leftrightarrow 2$ notation denotes an expression constructed by applying the exchange operator to the preceding terms, which ensures the wave function satisfies the exchange symmetry requirements dictated by particle indistinguishability.
This kind of entangled Gaussian state is practical for implementation in quantum technologies because such states can be readily produced and reliably controlled \cite{Georgiev2021,bergschneider2019experimental,ayatollah2024non}. 

\begin{figure}[h]
\centering
\begin{tikzpicture}
\draw (0,0.0) node[above right]{\includegraphics[width=8.5cm]{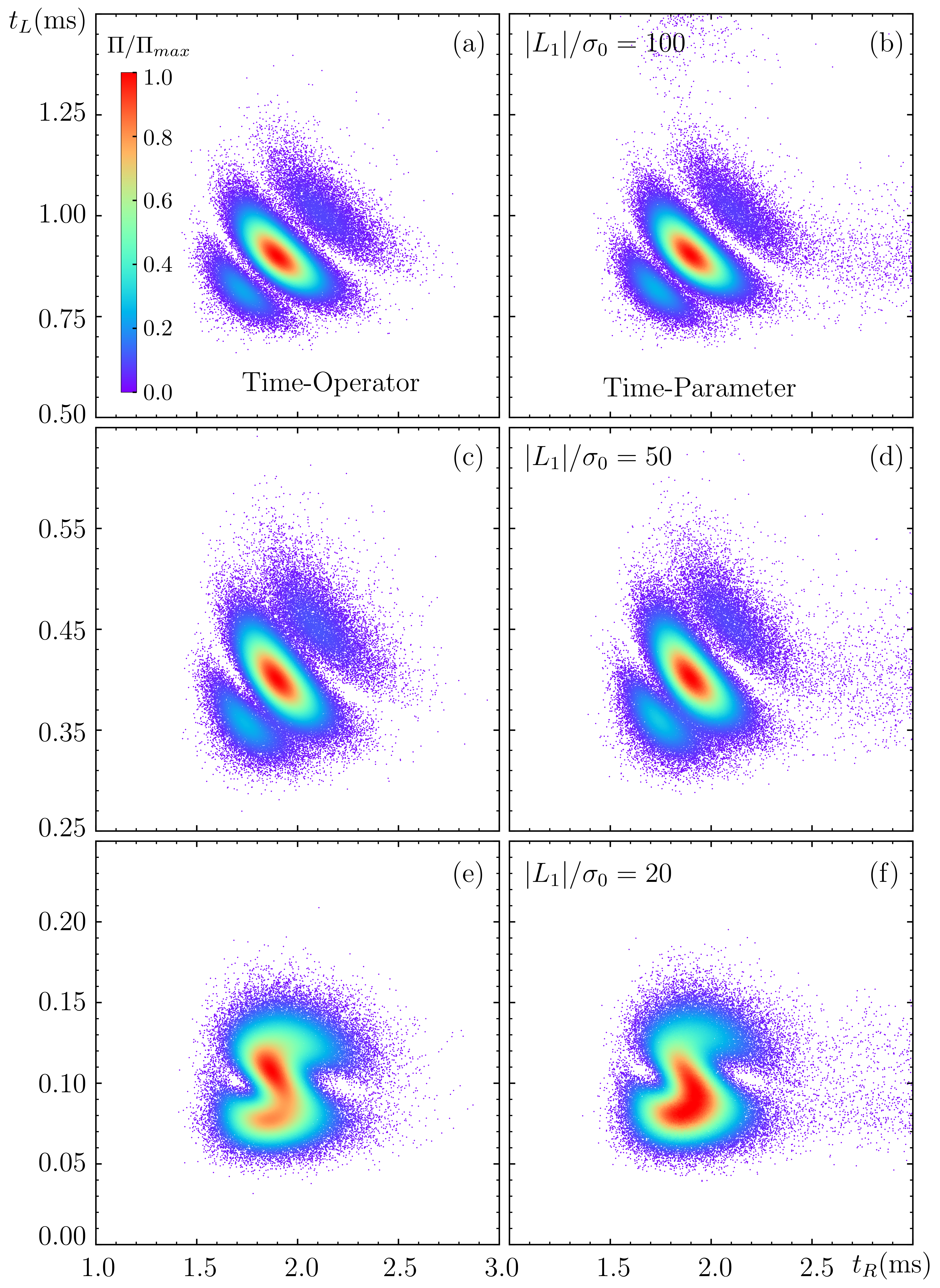}};
\end{tikzpicture}
\caption{\textbf{Comparing arrival-time distributions in the near-field and far-field regimes.}
Panels (a), (c), and (e) (left column) represent two-particle arrival-time distributions resulted from the time-operator approach, whereas panels (b), (d), and (f) (right column) display the corresponding distributions computed using the time-parameter approach. The rows indicate different left-detector positions, arranged from the far-field at the top to the near-field regime at the bottom. As seen, the two approaches differ more clearly in the near-field regime.
Each scatter plot is generated using $10^5$ sample points. The parameters of the initial wave function are chosen as $\sigma_0=10^{-6}$m, $u=10^{-1}$m/s, $l_A=8\sigma_0$, and $l_B=12\sigma_0$. The right detector is placed at $X_R=200\sigma_0$. In panels (a)-(b), (c)-(d), and (e)-(f), the left detector is placed at $X_L=-100\sigma_0$, $X_L=-50\sigma_0$, and $X_L=-20\sigma_0$, respectively. All the distributions are computed for Helium atoms with mass $m=6.64\times10^{-27}$kg. In Each panel, the colormap is normalized to $\Pi_{max}$, the maximum of arrival time probability density. In the right-column plots, the detection temporal resolution is fixed to be $\tau=10^{-6}$s.
}\label{fig2}
\end{figure}

Using Eq.~(\ref{OTA-P-rep}) together with Eq.~(\ref{Initial-wave-function}), we plot in the left column of Fig.~\ref{fig2} the joint arrival-time probability density within the time-operator framework. In this figure, the scatter plots are generated using $10^5$ sample points, and the parameters have been chosen as $\sigma_x=10^{-6}$(m), $u_x=0.1$(m/s), $l_A = 8\times\sigma_0$(m), $l_B = 12\times\sigma_0$(m)  and $m_1 = m_2 = 6.646 \times 10^{-27}$ (kg). These values are in agreement with the proposed setup in reference \cite{kofler2012einstein} in which colliding Helium-4 atoms have been used for producing the initial entangled state \cite{perrin2007observation}---The double-double-slit setup introduced in the reference \cite{kofler2012einstein} can be utilized to produce our initial entangled wave function. Furthermore, analogous initial entangled states can be realised and precisely controlled in optical trapping configurations, providing an alternative experimental platform in which the degree of entanglement can be systematically engineered and the arrival-time distribution can be accurately measured \cite{bergschneider2019experimental}.

In order to compute the arrival-time distribution within the time-parameter framework, one must first determine the time evolution of the initial wave function (\ref{Initial-wave-function}), taking into account the detector’s back-effect. Since the free two-particle Hamiltonian is separable, the time evolution of this wave function can be found from the time evolution of each single-particle wave packet as follows:
\begin{eqnarray}\label{Psi-k}
\!\!\Psi_{k}(x_1,x_2)\! &=& \! N_k\!\left( A_k^+ (x_1)A_k^-(x_2)\!+\! B_k^+(x_1)B_k^-(x_2) \right. \ \ \ \ \nonumber \\ 
\! &+& \left.\!  1  \leftrightarrow  2 \right)\!,
\end{eqnarray}
in which $N_k$ is the normalization constant, $A_k^{\pm}(x)$ and $B_k^{\pm}(x)$ are constructed from the time evolution of the associated initial Gaussian wave functions---according to Eq.~(\ref{one-particle-evolution})---as follows:
\begin{eqnarray}
A_k^{\pm}(x)&\!\!=\!\!&N_k^{A^\pm}\langle x|e^{-i\hat{H}\tau/\hbar} \left(\hat{\bar{P}}e^{-i\hat{H}\tau/\hbar}\right)^{k-1}|A^\pm_{0}\rangle, \ \ \ \ \\
B_k^{\pm}(x)&\!\!=\!\!&N_k^{B^\pm}\langle x|e^{-i\hat{H}\tau/\hbar} \left(\hat{\bar{P}}e^{-i\hat{H}\tau/\hbar}\right)^{k-1}|B^\pm_{0}\rangle, \ \ \ \
\end{eqnarray}
where, $N_k^{A^\pm}$ and $N_k^{B^\pm}$ represent normalization constants and, $|A^\pm_{0}\rangle$ and $|B^\pm_{0}\rangle$ are vector states associated with the initial wave packets $A_0^{\pm}(x)$ and $B_0^{\pm}(x)$, respectively. 
Moreover, according to Eq~(\ref{psi-ket-R}) and Eq~(\ref{psi-ket-L}), the single-particle wave functions are given by
\begin{eqnarray}
\psi^{(L)}_{k,l}(x)\!\! &=&\!\! N^{(L)}_{kl}\Big(\alpha_{k,R}^{-} A^+_{k+l}(x) +\beta_{k,R}^{-} B^+_{k+l}(x) \nonumber \\
 & + &  \alpha_{k,R}^{+} A^-_{k+l}(x) +\beta_{k,R}^{+} B^-_{k+l}(x) \Big)
\label{psi-k1}
\end{eqnarray}
\begin{eqnarray}
\psi^{(R)}_{k,l}(x)\!\! &=&\!\! N^{(R)}_{kl}\Big(\alpha_{k,L}^{-} A^+_{k+l}(x) +\beta_{k,L}^{-} B^+_{k+l}(x) \nonumber \\
 & + &  \alpha_{k,L}^{+} A^-_{k+l}(x) +\beta_{k,L}^{+} B^-_{k+l}(x) \Big) \label{psi-k2}
\end{eqnarray}
where $\alpha_{k,L}^{\pm}=A^{\pm}_k(X_L)$, $\alpha_{k,R}^{\pm}= A^{\pm}_k(X_R)$, $\beta_{k,L}^{\pm}=B^{\pm}_k(X_L)$, and $\beta_{k,R}^{\pm}=B^{\pm}_k(X_R)$.   
Using the wave functions (\ref{Psi-k}), (\ref{psi-k1}), and (\ref{psi-k2}), one can numerically compute  the joint arrival-time probability density in the time-parameter approach, which is given in Eq.(\ref{Joint-TPA}). 
It is worth noting that the numerical simulation of Eq.~(\ref{Joint-TPA}) is substantially more demanding than the corresponding single-particle case, Eq.~(\ref{one-part-time-param}). The increased computational complexity arises not only from the higher dimensionality of the problem, but also from the larger number of integrals that must be evaluated—particularly when $\tau$ is small compared to the simulation time $T$.

\begin{figure*}[t]
\centering
\begin{tikzpicture}
\draw (-1,0.0) node[above right]{\includegraphics[width=18cm]{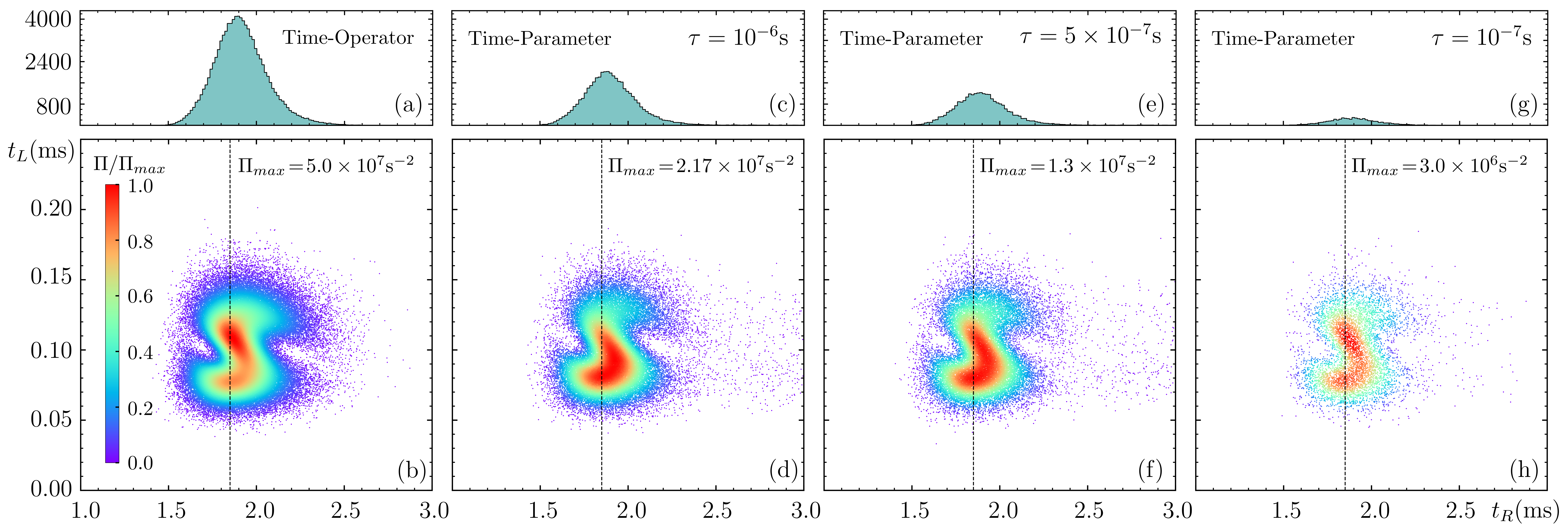}};
\end{tikzpicture}
\caption{\textbf{Detector temporal resolution effect on the arrival-time distribution.}
Panel (b) shows the two-particle arrival-time distribution obtained from the arrival time–operator approach, with its corresponding marginal shown in panel (a). Panels (d), (f), and (h) show the two-particle arrival-time distributions obtained from the time–parameter approach, with detector temporal resolutions $\tau = 10^{-6}\mathrm{s}$, $\tau = 5 \times 10^{-7}\mathrm{s}$, and $\tau = 10^{-7}\mathrm{s}$, respectively---their corresponding marginals appear in panels (c), (e), and (g).
In all panels, the right detector is placed at $X_R = 200\sigma_0$ and the left detector at $X_L = -20\sigma_0$. The distributions are computed for Helium atoms using the same initial wave-function parameters as in Fig.~(\ref{fig2}).
All colormaps are normalized to the maximum arrival-time probability density of the time-operator distribution. In the time-parameter approach, higher temporal resolution lowers the total detection probability due to the quantum Zeno effect, reflected in the fewer sampled points in panels (d), (f), and (h).}\label{fig3}
\end{figure*}

To address these challenges, we developed a computationally efficient Python implementation of the Crank-Nicolson method for solving the Schr\"odinger equation, as well as a Simpson-rule integration scheme, using Numba-compatible functions \cite{lam2015numba} to speedup the numerical computations.
As a result, in the right column of the Fig.~\ref{fig2}, the joint arrival-time distribution Eq.~(\ref{Joint-TPA}), is plotted.  In this column, the temporal detection resolution is chosen as $\tau=10^{-6}$s, and the parameters of the initial wave function are the same as one described earlier for the left-column. Moreover, in all panels of this figure, the position of the right detector is fixed at  $X_R=200\sigma_0$, while the position of the left detector varies form the far-field to the near-field regime---In panels (a)-(b), (c)-(d), and (e)-(f), the left detector is placed at $X_L=-100\sigma_0$, $X_L=-50\sigma_0$, and $X_L=-20\sigma_0$, respectively.

The numerical results allow for a detailed comparison of the two competing formulations of arrival-time in quantum mechanics. Interestingly enough, although these approaches are fundamentally different, they yield nearly identical distributions in the far-field regime, as seen in the top panels of Fig.~\ref{fig2}.
In this regime, both approaches predict non-classical \textit{multi-particle interference in arrival time}, a quantum phenomenon that has recently been predicted for atoms \cite{ayatollah2024non} and observed using photons \cite{ono2024quantum,jin2015spectrally,gerrits2015spectral,yu2025exploiting} in conceptually similar setups.

In contrast, in the near-field regime—shown in the bottom panels of Fig.~\ref{fig2}—the shapes of the resulting distributions differ noticeably, and thus the distinction between the two approaches becomes more pronounced. For example, the two approaches yield different mean arrival times in this regime. Consequently, this regime provides a promising setting for performing measurements that could determine which aforementioned arrival-time framework, if any, is compatible with experimental data. Moreover, even in the far-field regime, the arrival-time distributions obtained from the time-parameter approach exhibit small tails, which can be used to experimentally distinguish between the two approaches. However, observing this feature is experimentally challenging due to its very small magnitude.

\begin{figure}[h]
\centering
\begin{tikzpicture}
\draw (0,0.0) node[above right]{\includegraphics[width=8.5cm]{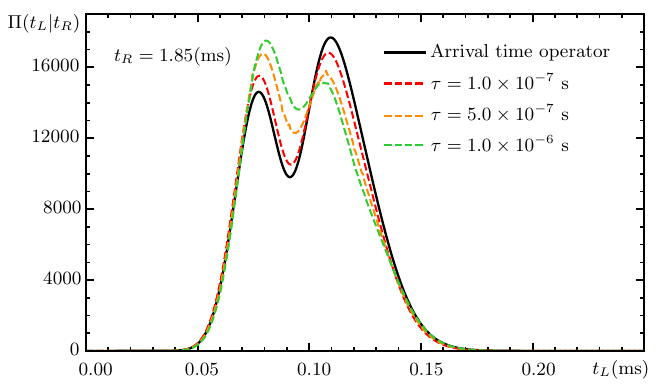}};
\end{tikzpicture}
\caption{\textbf{Conditional arrival-time probability density.} 
The plot shows the \textit{normalized} conditional arrival-time probability densities for the left detector, $\Pi(t_L|t_R)$, evaluated at the fixed right detection-time $t_R = 1.85\,\mathrm{ms}$ (corresponding to the vertical lines in Fig.~\ref{fig3}). The solid curve depicts the  arrival-time probability density obtained from the time-operator approach, while the dashed curves represent the results from the time-parameter approach. As seen, the results of two approaches coincide, when $\tau$ decrease.}\label{fig4}
\end{figure}
Understanding how the arrival-time distribution in the time-parameter approach depends on the detection temporal resolution is essential for optimizing experimental settings to enable a meaningful comparison between the two approaches. In this regard, in Fig.~\ref{fig3}, we plot the arrival-time distribution for several values of the detection temporal resolution, where other parameter values are chosen the same as in the bottom panels of Fig.~\ref{fig2}. As expected from the quantum Zeno effect, increasing the detection frequency suppresses the total detection probability, a behavior reflected in the reduced number of recorded detection events as $\tau$ decreases---see also the top panels of Fig.~\ref{fig3}. This result is in agreement with previous studies that numerically investigate the time-parameter approach with finite $\tau$ for a single particle in a 1D lattice \cite{dhar2015quantum,thiel2018first}. Furthermore, Fig.~\ref{fig3} indicates that although the total detection probability decreases, the difference between the shapes of the distributions obtained from the two arrival-time approaches becomes less pronounced as the detection resolution increases. For a more detailed examination, Fig.~\ref{fig4} presents the conditional arrival-time distribution at the left detector, $\Pi(t_L|t_R)$, corresponding to a slice of the joint distributions in Fig.~\ref{fig3} indicated by vertical lines. This figure demonstrates that decreasing $\tau$ reduces the discrepancy between the predictions of the time-parameter and time-operator approaches, suggesting that very fine temporal monitoring may diminish the distinguishability of the two frameworks.

The simulation results presented in Figs. \ref{fig2}, \ref{fig3}  and \ref{fig4} clearly demonstrate that a detector with microsecond temporal resolution---such as those reported in Refs. \cite{kurtsiefer19962A,kurtsiefer1997measurement}---is fully sufficient to unambiguously discriminate between the two approaches. Given that the characteristic temporal scale of the distributions is on the order of $100\mu$s, such resolution enables accurate reconstruction of the full temporal profiles.
Finally, it is important to remark that the preparation of the required initial wave function is readily achievable with existing atom-optics technologies, including established methods for the generation, optical cooling, and manipulation of entangled atoms \cite{cronin2009optics,lopes2015atomic,bergschneider2019experimental,anders2021momentum}.


\section{Summary and out-look}\label{Summary}
In this work, we extend two distinct treatments of the arrival-time problem within standard quantum theory to multi-particle systems and propose an experimentally feasible scheme that can discriminate between their predictions. Numerical simulations performed across a broad region of parameter space show that, although the two approaches generally yield nearly identical results—most notably in the far-field regime—there exists an experimentally accessible domain in which their predictions diverge substantially. This deviation is sufficiently large to make the two formulations experimentally distinguishable using current atom-interferometry techniques. Moreover, by examining the dependence of the arrival-time distributions on the detection resolution, we find that, although the time-parameter approach exhibits a suppressed detection probability near the Zeno regime, the discrepancy between the \textit{shapes} of the arrival-time distributions predicted by the two approaches nevertheless decreases as~$\tau$ is reduced.

Beyond the fundamental questions addressed here, our results also have direct implications for practical applications in atom optics that rely on entangled particles, including atomic ghost imaging \cite{khakimov2016ghost,ryczkowski2016ghost}, entanglement-enhanced gravimetry \cite{geiger2018proposal}, and related quantum-sensing technologies \cite{brown2023time,tenart2021observation}. A detailed analysis of these applications is left for future work. Nevertheless, it is important to emphasize that realistic implementations will require extending the present studies to interacting multi-particle systems subject to external gravitational fields. Such extensions are computationally demanding and will require the development of optimized numerical methods, thereby defining an important direction for future research.

Moreover, at least for applications in quantum optics, it is essential to extend these approaches for relativistic particles, particularly photons. This represents an important research direction, as the measurement of photon arrival-time distributions using state-of-the-art ultrafast detectors \cite{korzh2020demonstration,abazi2025multiphoton,kawasaki2025real} forms a cornerstone of modern quantum technologies---for example see \cite{xavier2025energy,yu2025exploiting,roeder2024measurement}. At a fundamental level, however, this remains a challenging problem. While a natural relativistic generalization of the Aharonov–Bohm time-of-arrival operator has been proposed \cite{flores2022relativistic,razavy1969quantum}, it is defined in terms of a position operator, whose definition in a relativistic context remains controversial \cite{terno2014localization,rosenstein1987explicit,newton1949localized}\footnote{\textcolor{blue}{However, Kijowski has derived a similar relativistic arrival-time distribution via an axiomatic approach, without directly using a position operator \cite{kijowski1974time}. Nevertheless, the validity of this derivation remains a matter of debate \cite{das2021times}.}}.
Similarly, the time-parameter approach relies on a well-defined position probability density, which is also problematic in the relativistic regime and lacks a universally accepted definition \cite{kowalski2011salpeter,rosenstein1985probability,hegerfeldt1985violation,wagner2012local,sebens2019electromagnetism},  at least for bosons. Recently, some of the present authors have developed a framework for quantum mechanics that does not rely on a self-adjoint position operator \cite{kazemi2025non,kazemi2018probability}, offering a promising avenue to address these issues. Detailed investigations of relativistic extensions and their implications for photon-based quantum technologies are left for future works.

Finally, it is important to remark that in principle it is  possible that, in a sufficiently accurate experiment, neither mention approach will match the measured data, indicating the need for a more realistic modeling of the detection process. This possibility may, in turn, bring us closer to an experimental investigation of the \textit{measurement problem} \cite{maudlin1995three}.
In this regard, comparing our results with arrival-time predictions in alternative quantum theories—particularly continuous-collapse models \cite{ghirardi1986unified,bassi2013models} and Bohmian mechanics \cite{bohm1952suggested,ayatollah2024non}—constitutes an important direction for future work.

\section*{Acknowledgement}
The authors acknowledge access to the high-performance computing cluster at  University of Qom, which made the numerical simulations in this work possible. SSH is supported in part by the by the “Quantum Technologies for Sustainable Development” project from the National Institute for Theoretical and Computational Sciences (NITheCS). 
M.K would like to thank Mohammad Khorrami for insightful discussion.


\bibliography{bibliography}
\end{document}